\documentclass[superscriptaddress,twocolumn, showpacs,nofootinbib,longbibliography,notitlepage]{revtex4-2}
\usepackage{etex}
\usepackage{physics}
\usepackage{amsmath,amssymb,amsthm,amstext,mathrsfs,bm}
\usepackage[pdftex]{graphicx}
\usepackage{times,txfonts}
\usepackage{braket}
\usepackage{color}
\usepackage{natbib}
\usepackage{amsmath,blkarray}
\usepackage{mathtools}
\usepackage{latexsym}
\usepackage{tabularx, booktabs}
\usepackage{graphics,epstopdf}
\usepackage{graphicx}
\usepackage{float}
\usepackage{graphicx}
\usepackage{amsfonts}
\usepackage{subcaption}
\usepackage{color,soul}
\usepackage[inkscapelatex=true]{svg}
\usepackage{hyperref}
\usepackage{placeins}

\newtheorem{thm}{Theorem}

\begin{document}
	
	\title{Noise robust super-Heisenberg scaling in time-dependent quantum sensing by subspace-twirled dynamical decoupling}

\author{Bakmou Lahcen}
\affiliation{ Institute of Fundamental and Frontier Sciences,
University of Electronic Science and Technology of China, Chengdu 610051, China}
\affiliation{Key Laboratory of Quantum Physics and Photonic Quantum Information, Ministry of Education, University of Electronic Science and Technology of China, Chengdu 611731, China}

\author{Ke Zeng}
\affiliation{ Institute of Fundamental and Frontier Sciences,
University of Electronic Science and Technology of China, Chengdu 610051, China}
\affiliation{Key Laboratory of Quantum Physics and Photonic Quantum Information, Ministry of Education, University of Electronic Science and Technology of China, Chengdu 611731, China}

\author{Tianyu Yang}
\affiliation{ Institute of Fundamental and Frontier Sciences,
University of Electronic Science and Technology of China, Chengdu 610051, China}
\affiliation{Key Laboratory of Quantum Physics and Photonic Quantum Information, Ministry of Education, University of Electronic Science and Technology of China, Chengdu 611731, China}

\author{Yu Jiang}
\affiliation{ Institute of Fundamental and Frontier Sciences,
University of Electronic Science and Technology of China, Chengdu 610051, China}
\affiliation{Key Laboratory of Quantum Physics and Photonic Quantum Information, Ministry of Education, University of Electronic Science and Technology of China, Chengdu 611731, China}

\author{Kok Chuan Tan}
\affiliation{ Institute of Fundamental and Frontier Sciences, University of Electronic Science and Technology of China, Chengdu 610051, China}
\affiliation{Key Laboratory of Quantum Physics and Photonic Quantum Information, Ministry of Education, University of Electronic Science and Technology of China, Chengdu 611731, China}
\affiliation{School of Physics, University of Electronic Science and Technology of China, Chengdu 610054, China}
\affiliation{School of Science, Xihua University, Chengdu 610039, China}
\email{tankokchuanbobby@uestc.edu.cn}
	%%%%%%%%%%%%%%%%%%%%%%%%%%%%%%%%%%%%%%%%%%%%%%%%%%%%%%%%%%%%%%%%%%%%%%%%%%%%%%%%%%%%%%%%%%%%%%%%%%%%%%%%%%%%%%%%%%%%%%%%%%%%%%%%%%%%%%%%%%%%%%%%%%%%%%%%%%%%%%%%%%%%%%%%%%%%%%%%%%%%%%%%%%%%%%%%%%%%%%%%%%%%%%%%%%%%%%%%%%%%%%%%%%%%%%%%%%%%%%%%%%%%%%%%%%%
	\begin{abstract}
		In quantum parameter estimation, time-dependent Hamiltonians can enable ultraprecise estimates of a physical parameter beyond the standard Heisenberg limit, also called the super-Heisenberg limit. However, achieving optimal scaling of the measurement precision typically requires an adaptive quantum control scheme which is resource-intensive and hard to implement in general. Furthermore, this advantage is often severely eroded by environmental noise. We propose the use of a dynamical decoupling scheme based on subspace twirling -- the averaging of random unitaries over a subspace rather than the entire Hilbert space -- to overcome these two limitations. We first prove necessary and sufficient conditions for subspace twirling to preserve the encoding signal while simultaneously removing environmental noise. We then establish conditions and provide performance guarantees for when time dependent Hamiltonians undergoing subspace twirling can achieve super-Heisenberg scaling without adaptive controls. Finally, we illustrate our results using an explicit example where a qubit sensor undergoing a subspace twirl simultaneously suppresses noise and achieves super-Heisenberg scaling. These results offer guidance for the design of ultraprecise quantum sensors under time-dependent and noisy conditions.
	\end{abstract}
	\maketitle
	%%%%%%%%%%%%%%%%%%%%%%%%%%%%%%%%%%%%%%%%%%%%%%%%%%%%%%%%%%%%%%%%%%%%%%%%%%%%%%%%%%%%%%%%%%%%%%%%%%%%%%%%%%%%%%%%%%%%%%%%%%%%%%%%%%%%%%%%%%%%%%%%%%%%%%%%%%%%%%%%%%%%%%%%%%%%%%%%%%%%%%%%%%%%%%%%%%%%%%%%%%%%%%%%%%%%%%%%%%%%%%%%%%%%%%%%%%%%%%%%%%%%%%%%%%%
	\textit{Introduction}---Ultraprecise measurement devices are central to many modern quantum technologies, including gravitational-wave detectors~\cite{schnabel2010quantum}, optical atomic clocks~\cite{kruse2016improvement}, and quantum sensors~\cite{pirandola2018advances, thomas2011real}. Quantum metrology seeks to surpass classical precision limits by exploiting  quantum effects such as coherence, entanglement, and coherent control to optimize parameter estimation~\cite{giovannetti2004quantum, giovannetti2011advances, DEMKOWICZDOBRZANSKI2015345, pezze2018quantum, yuan2016sequential, liu2017quantum}. In the quantum regime, the precision limits of a single parameter estimation task is fundamentally bounded by the Quantum Cramér--Rao Bound (QCRB), which is in turn determined by the Quantum Fisher Information (QFI). For linear parameter encoding under time-independent Hamiltonians, coherent evolution implies that the QFI growing quadratically as $\mathcal{F}\sim t^2$~\cite{giovannetti2006quantum, demkowicz2012elusive} where $t$ is the interaction time, which is conventionally referred to as Heisenberg scaling (HS). Remarkably, this limit is not fundamental when the parameter encoding is time-dependent and/or nonlinear, and the QFI can scale as $\mathcal{F}\sim t^4$, or more generally as $t^{2m+2}$ where $m \geq 0$. This is known as super-Heisenberg scaling (SHS) \cite{pang2017optimal,gefen2017control,gefen2019overcoming,rams2018limits,roy2008exponentially,yousefjani2025nonlinearity,sarkar2025exponentially, dey2025quantum}. Such enhanced scalings can arise from time-dependent interference and nonlinear dynamical effects, offering the prospect of precision far beyond conventional HS.
	
	In practice, protocols achieving the optimal SHS require implementing adaptive quantum controls \cite{pang2017optimal}. The maximal QFI for time-dependent Hamiltonians is generally achieved by interleaving the sensing evolution with real-time measurements and feedback that continuously updates the control fields. Such adaptive schemes are experimentally expensive, requiring rapid reconfiguration of the control fields and interrogation times much faster than the decoherence time of the system ~\cite{Naghiloo2017}.
	
	On the subject of decoherence, practical quantum sensors will inevitably interact with the environment, resulting in noise. Environmental noise rapidly erodes the quantum advantage by destroying the coherent interference responsible for quantum effects. Methods such as Quantum Zeno Dynamics (QZD), Dynamical Decoupling (DD) and Quantum Error Correction (QEC) have demonstrated that HS can be recovered in noisy systems under suitable conditions~\cite{dur2014improved,unden2016quantum,layden2019ancilla,zhou2018achieving,zeng2026achievingheisenberglimitnoisy}. These results typically make simplifying assumptions to make the problem more tractable, such as Markovian noise and/or time-independent interactions. Consequently, it is desirable to find a protocol that can achieve SHS under more general assumptions about the noise.
	
	In this work, we propose a dynamical decoupling scheme based on subspace twirling to achieve SHS in time-dependent quantum metrology under noisy conditions. In conventional DD, the system-environment dynamics are completely frozen by twirling over the entire Hilbert space $\mathcal{H}$. In subspace twirling, we identify a strict subspace of $\mathcal{H}$ and average over random unitary operations acting on this subspace. While twirling can always suppress environmental noise, the encoding of a physical parameter $\theta$ may not necessarily survive the twirling process, in which case it is no longer useful for quantum metrology. To this end, we prove necessary and sufficient conditions for subspace twirling to achieve noise suppression while preserving the encoding of parameter $\theta$. We also establish conditions where subspace twirling can allow SHS to be achieved without requiring adaptive controls. The combination of these results imply that subspace twirling can simultaneously (i) suppress time-dependent noise, and (ii) remove the requirement for adaptive controls while reaching SHS. Finally, we demonstrate our findings by discussing an explicit example where a qubit sensor in a rotating field is able to suppress noise and achieve SHS simultaneously by twirling over a single subspace.
	
	\textit{Preliminaries}---In the subsequent sections, we will assume idealized,  instantaneous, error-free control pulses, also called bang-bang operations, are available \cite{Viola1998}.
	
	Let $\mathcal{H}_S$ be a finite-dimensional Hilbert space of dimension $d$. Consider a single parameter estimation setting where the probe state in $\mathcal{H}_S$ evolves under a general time-dependent Hamiltonian of the following form: 
	\begin{equation} \label{eq::H_tot}
		H_{\rm {tot}}(\theta, t) = H_{\rm S}(\theta, t) \otimes \openone_{\rm B} + \openone_{\rm S} \otimes H_{\rm B}(t) + H_{\rm SB}(t),
	\end{equation}
	where $H_{\rm S}(\theta,t) = \sum\nolimits_{i=1}^{r} g_i(\theta,t)\, G_i$ is the signal Hamiltonian encoding the parameter $\theta$, $G_i$ are Hermitian operators, $g_i(\theta,t)$ are time-dependent functions encoding unknown parameter $\theta$, and $H_{\rm SB}(t) = \sum_j  h_j(t) \, S_{j}\otimes B_{j}$ is a generic time dependent, system-bath interaction in some operator basis $\{S_i\}$ and $\{ B_j \}$. Note that the encoding of parameter $\theta$ is confined within the subsystem $S$, and that all system-bath interactions satisfying this requirement can always be written in the form of Eq.~\ref{eq::H_tot}.
	
	Suppose an initial probe state $\ket{\psi_0}$ evolves under a unitary $U(\theta,t)=\mathscr{T}\exp{-i\int_0^t H\qty(\theta, s) ds}$ generated by a time and parameter dependent Hamiltonian $H(\theta,t)$, where $\mathscr{T}$ is the time ordering operator. After interaction time $t$, the instantaneous state is  $\ket{\psi(\theta,t)} = U(\theta,t)\ket{\psi_0}$, where $\ket{\psi_0}$ is the initial probe state. For a given probe state and unitary evolution, the QFI obeys $\mathcal F(\theta) = 4\,\operatorname{Var}[h(\theta,t)]_{\ket{\psi(0)}}$, where $h(\theta,t) = i\,U^\dagger(\theta,t)\partial_\theta U(\theta,t)$ is the generator, and the variance is evaluated with respect to the state $\ket{\psi(0)}$ \cite{pang2014quantum}. When maximized over all possible $\ket{\psi_0}$, the  QFI is  upper bounded by  \begin{align} \label{eq::upperBound} \max_{\ket{\psi_0}} \mathcal F(\theta) \le \left[ \int_0^t \left( \lambda_{\max}(s) - \lambda_{\min}(s) \right) ds \right]^2,\end{align} where $\lambda_{\max}(t)$ ($\lambda_{\min}(t)$) are the maximum (minimum) eigenvalues of $\partial_\theta H(\theta,t)$, with corresponding eigenvectors$\ket{\lambda_{\max}(t)}$ ($\ket{\lambda_{\min}(t)}$). In particular, this bound can be saturated when: \begin{itemize}
		\item[(A)] $\ket{\psi_0} = \left (\ket{\lambda_{\max}(0)} \pm \ket{\lambda_{\min}(0)} \right)/\sqrt{2}$ is the initial probe state;
		\item[(B)] $U(\theta,t)\ket{\lambda_{\max}(0)}$ ($U(\theta,t)\ket{\lambda_{\min}(0)}$) are eigenvectors corresponding to the maximum (minimum) eigenvalues of $\partial_\theta H(\theta,t)$ throughout the evolution up to time $t$.
	\end{itemize} An adaptive control \cite{pang2017optimal} is required to saturate this bound in general. However, if $U(\theta,t)$ commutes with $\partial_\theta H(\theta,t)$ for all $t$, and there are no level crossings between the maximum and minimum eigenvalues, the bound can be saturated with no additional controls under noiseless conditions.
	
    In DD, a time-dependent control term $H_c(t)$ is added to an initial Hamiltonian $H_0(t)$ such that $H(t) = H_0(t)+H_c(t)$. By defining $U_c(t) \coloneqq \mathscr{T} \exp \{ -i \int_{0	}^{t} H_c(s) \, ds \}$, the effective Hamiltonian in the toggling frame is given by $\tilde{H}(t) \coloneqq U_c(t)^\dagger H_0(t) U_c(t)$. The goal of DD is to design the control $H_c(t)$ such that subsystem $S$ is decoupled from the bath $B$ in the effective Hamiltonian $\tilde{H}(t)$. Notably, in Refs.~\cite{Viola1999,Viola2003} it was shown that given a finite unitary group $\mathcal{G}$, and assuming the bang-bang control limit, one can achieve the effective Hamiltonian \begin{align}
    	\tilde{H}(t) &\approx \frac{1}{t} \int_{0}^{t} U_c(s)^\dagger H_0(s) U_c(s) \, ds  \label{eq::firstOrder}\\
    	&= \frac{1}{\abs{\mathcal{G}}} \sum_{U_g \in \mathcal{G}} U_g^\dagger H_0(t) U_g \\
    	&= \mathcal{T}_\mathcal{G}[H_0(t)] \label{eq::twirl},
    \end{align} where we define the twirling over the unitary group $\mathcal{G}$ as the linear map $\mathcal{T}_G[H] \coloneqq \frac{1}{\abs{\mathcal{G}}} \sum_{U_g \in G} U_g^\dagger H U_g.$ Note that Eq.~\ref{eq::firstOrder} is the first order approximation in the Magnus expansion of $\tilde{H}(t)$.
    
    In this work, we will consider twirling operations where group $\mathcal{G}$ is the set of all possible unitary operators acting on a strict subspace $\mathcal{V} \subset \mathcal{H_S}$. Since this group is defined completely by the subspace $\mathcal{V}$, we will denote the twirling map as $\mathcal{T}_{\mathcal{V}}$. Strictly speaking, $\mathcal{T}_{\mathcal{V}}$ is a twirling operation over a continuous group, whereas a finite group twirling was considered in Eq.~\ref{eq::twirl}. However, for any finite dimensional subspace $\mathcal{V}$ there exists a finite set of unitaries $\{U_i\}_{i=1}^{q}$ called a unitary 1-design such that $\mathcal{T}_{\mathcal{V}} (H) = \frac{1}{q}\sum_{i=1}^q  U_i H U_i^\dagger$ \cite{Dankert2009}. This allows Eq.~\ref{eq::twirl} to be applied.
    
    \textit{Suppressing noise with subspace twirling}--- Consider the total Hamiltonian $H_{\rm tot}(t)$ in Eq.~\ref{eq::H_tot}. The noise is contributed by the system bath term $H_{\rm SB}(t) = \sum_j  h_j(t) \, S_{j}\otimes B_{j}$, while the parameter $\theta$ to be estimated is encoded by the term $H_{\rm S}(\theta,t) = \sum\nolimits_{i=1}^{r} g_i(\theta,t)\, G_i.$ We can assume that $S_i$ are traceless Hermitian operators since any component with nonzero trace contributes a term proportional to $\openone \otimes B_j$ which can be grouped with the $\openone_S \otimes H_B(t)$ in Eq.~\ref{eq::H_tot}. 
    
    We now consider the problem of suppressing time dependent noise, which is one of two main goals identified in the Introduction. 
    To successfully decouple the system-bath interaction, we need to find subspace $\mathcal{V} \subset \mathcal{H}_S$ such that 
    \begin{align} \label{eq::decouple}
    		\mathcal{T}_{\mathcal{V}} \otimes \openone [H_{\rm SB}(t)] &= \sum_j h_j(t)\,\mathcal{T}_{\mathcal{V}}(S_i) \otimes B_i \notag \\
    		& \propto \openone \otimes B'(t).
    \end{align}
    
   As discussed in the Preliminaries, it is always possible to satisfy this condition by twirling, for instance, by choosing $\mathcal{V}$ to be the entire Hilbert space $\mathcal{H}_S$. However, this will also remove $H_S(\theta,t)$ which encodes the parameter $\theta$. For twirling to be useful in quantum metrology, we need to impose at least two extra conditions, which are \begin{align} \label{eq::signalCondition1}
   	\mathcal{T}_{\mathcal{V}} [H_S(\theta,t)] \not\propto H_S'(t),
   \end{align} and \begin{align} \label{eq::signalCondition2}
   \mathcal{T}_{\mathcal{V}} [H_S(\theta,t)]  \not\propto \openone,
   \end{align} where $H_S'(t)$ is some Hermitian operator that is independent of the parameter $\theta$. In other words, after subspace twirling, the effective Hamiltonian must still encode information about $\theta$ (hence $\not\propto H_S'(t)$) and generate nontrivial evolution (hence $\not\propto \openone$). Successfully suppressing noise while still being useful for metrology is therefore equivalent to finding a strict subspace $\mathcal{V} \subset \mathcal{H}_S$ that satisfies Eqs.~\ref{eq::decouple}, \ref{eq::signalCondition1} and~\ref{eq::signalCondition2}.
   
   Our first main result establishes when the conditions above can be satisfied. It is  stated as the following Theorem:
   \begin{thm} \label{thm::main1}
   	Given $H_{\rm tot}(\theta, t)$, there exists a strict subspace $\mathcal{V}$ and corresponding subspace twirling $\mathcal{T}_{\mathcal{V}}$ that satisfies the noise suppression and signal preserving conditions  Eqs.~\ref{eq::decouple}, \ref{eq::signalCondition1} and~\ref{eq::signalCondition2} if and only if there exists a vector $\ket{v} \in \mathcal{H}_S$ such that (a) $\braket{v | S_i |v} = 0$ for every $i$ and (b) $\nu(\theta) \coloneqq \braket{v | H_S(\theta, t) |v} $ is not a constant function with respect to $\theta$.
   \end{thm}
   
   A full discussion and proof of Theorem~\ref{thm::main1} is provided in Ref.~\cite{SuppMa}. The above result simplifies the problem of finding a subspace twirling operator $\mathcal{T}_{\mathcal{V}}$  that satisfies all the required conditions and reduces it to finding a single vector $\ket{v}$ that satisfies conditions (a) and (b) in Theorem~\ref{thm::main1}. Furthermore, given any $\ket{v}$, one may efficiently check conditions whether (a) and (b) are true. This provides a useful method for verifying whether subspace twirled dynamical decoupling is useful for metrology. 

Note that Eq.~\ref{eq::decouple} implies full decoupling of the system-bath dynamics. When this condition is satisfied, it means that one can use \emph{any} initial probe state $\ket{\psi_0}$ and the subsequent dynamics will be noiseless. However it is often the case that one prepares a specific probe state, and all that is required is that this particular probe state's dynamics is not affected by noise. Suppose $\mathcal{W} = \operatorname{span} \{ \ket{w_i} \}_{i=1}^r$ is a subspace of $\mathcal{H}_S$, and one prepares some superposition of states $\ket{\psi_0} = \sum_{i=1}^r a_i \ket{w_i} \in \mathcal{W}$ from this subspace. For a probe state of this type, it suffices that noiseless dynamics can be enforced via subspace twirling within the subspace $\mathcal{W}$, so a full system-bath decoupling is not required. In this case, Theorem~\ref{thm::main1} can be applied by defining the projector $\Pi_{\mathcal{W}}$ onto the subspace $\mathcal{W}$ and then performing the replacements $ H_{\rm  SB}(t) \rightarrow ( \Pi_{\mathcal{W}}\otimes \openone) H_{\rm  SB} (t) (\Pi_{\mathcal{W}}\otimes \openone)$,  $H_S(\theta,t) \rightarrow \Pi_{\mathcal{W}} H_S(\theta,t) \Pi_{\mathcal{W}}$, and $\openone \rightarrow \Pi_{\mathcal{W}}$ in Eqs.~\ref{eq::decouple}, \ref{eq::signalCondition1} and~\ref{eq::signalCondition2}. The application of Theorem~\ref{thm::main1} in this scenario is discussed in greater detail in the Supplemental Material~\cite{SuppMa}.
   
\textit{Achievability of SHS without adaptive controls}---We now turn to the second goal of this work, which is to achieve HS/SHS without requiring adaptive controls. In the following, we say that HS/SHS is achieved if $\mathcal{F}(\theta) = \Omega(t^{l}) $ for some $l \geq 2 $, where $\Omega$ denotes the asymptotic lower bound. This means that for sufficiently large $t$, there exists a positive constant $c$ such that  $\mathcal{F}(\theta) \geq  c t^{l} $.

Consider a time-dependent signal Hamiltonian $H_{\rm S}(\theta,t) = \sum\nolimits_{i=1}^{r} g_i(\theta,t)\, G_i$. Without loss in generality, we can assume that $G_i$ are traceless, Hermitian operators. This is because any nonzero trace is contributed by a component proportional to $\openone$, which generates trivial dynamics. Recall from the Preliminaries that in order to saturate the upper bound in Eq.~\ref{eq::upperBound}, we must satisfy Conditions (A) and (B).

In order to satisfy Condition (B) without additional controls, we seek a twirled Hamiltonian $\mathcal{T}[H_S(\theta,t)]$ that satisfies the following: \begin{align} \label{eq::commuteCondition}
	\left[ \mathcal{T}[H_S(\theta,t)], \partial_\theta \mathcal{T}[H_S(\theta,t)] \right ] =0, \quad \forall t\geq 0
\end{align} where $[A,B] \coloneqq AB-BA$ is the commutator. If one allows for twirling over multiple subspaces, it is always possible to find a $\mathcal{T}$ that satisfies Eq.~\ref{eq::commuteCondition}. To see this, let $\mathcal{V}$ be any strict subset of the Hilbert space $\mathcal{H}_S$. Given $\mathcal{V}$, one can always find its orthogonal space $\mathcal{V}^\perp$. Let the twirling map $\mathcal{T} \coloneqq \mathcal{T}_{\mathcal{V}} \circ \mathcal{T}_{\mathcal{V}^\perp}$ be the twirling operator over both subspaces $\mathcal{V}$ and $\mathcal{V}^\perp$. One may then verify that $\mathcal{T}[H_S(\theta,t)]$ can be written in the following block matrix form: \begin{align}
\mathcal{T}[H_S(\theta,t)] = \begin{pmatrix}
	\frac{\mu(\theta,t)}{\Tr(\openone_{\mathcal{V}})} \openone_V & 0 \\
	0 & -\frac{\mu(\theta,t)}{\Tr(\openone_{\mathcal{V}^\perp})} \openone_{\mathcal{V}^\perp}
\end{pmatrix} ,
\end{align} where $\mu(\theta,t)$ is some function of $\theta,t$. Since $\mathcal{T}[H_S(\theta,t)]$ is diagonal, we immediately satisfy Eq.~\ref{eq::commuteCondition}. Note that satisfying Eq.~\ref{eq::commuteCondition} is only \emph{necessary} for Condition~(B), but not sufficient. If level crossings between the maximum and minimum occurs, then additional controls locally near the points where level crossings occur may be necessary \cite{pang2017optimal}. Even so, the maximal Fisher information $\max_{\ket{\psi_0}} \mathcal{F}(\theta)$ will depend on the function  $\partial_\theta\mu(\theta,t)$, so one expects that achieving SHS is possible when $\partial_\theta\mu(\theta,t)$ satisfies certain properties.

This leads to our second main result, which improves on the above argument by providing a checkable condition based on the operator norm $\Vert \cdot \Vert$ that is both necessary and sufficient.

\begin{thm} \label{thm::main2}
	Given $H_{\rm S}(\theta,t) = \sum\nolimits_{i=1}^{r} g_i(\theta,t)\, G_i$, there exists a strict subspace $\mathcal{V} \subset \mathcal{H}_S$ and corresponding twirling map $\mathcal{T} \coloneqq \mathcal{T}_{\mathcal{V}} \circ \mathcal{T}_{\mathcal{V}^\perp}$ such that $\mathcal{T}[H_S(\theta,t)]$ achieves HS or SHS without additional adaptive controls if and only if $\Vert K(\theta,t)\Vert = \Omega(t^k)$ for some $k \geq 1$ where $K(\theta,t) \coloneqq \int_{0}^{t} \partial_{\theta}H_S(\theta,s) \, ds $.
	
	Furthermore, if $\Vert K(\theta,t)\Vert = \Omega(t^k)$, then the QFI $\mathcal{F}(\theta) = \Omega(t^{2k})$ is achievable.
\end{thm}

A full proof of Theorem~\ref{thm::main2} can be found in Ref.~\cite{SuppMa}. Theorem~\ref{thm::main2} reduces the problem of optimizing over all possible twirling maps of the form $\mathcal{T} = \mathcal{T}_{\mathcal{V}} \circ \mathcal{T}_{\mathcal{V}^\perp}$ and reduces it to evaluating the operator norm of the matrix $K(\theta,t) \coloneqq \int_{0}^{t} \partial_{\theta}H_S(\theta,s) \, ds$. In quantum parameter estimation, it is assumed that the signal Hamiltonian $H_S(\theta,t)$ is known, while the parameter $\theta$ is not precisely known. Under this assumption, $K(\theta,t)$ and $\Vert K(\theta,t)\Vert$ can always be evaluated locally around any point $\theta$. Note that the operator norm can be efficiently calculated numerically via a semidefinite program \cite{Watrous2018}, and the scaling with time $t$ can be obtained from a logarithmic plot. This provides an efficient method to check whether HS/SHS is achievable using twirling map $\mathcal{T}$ alone. The result also gives guarantees on the achievable metrological performance, as $\Vert K(\theta,t)\Vert = \Omega(t^k)$ implies that a QFI of at least $\mathcal{F}(\theta) = \Omega(t^{2k})$ can be reached.

\textit{Qubit example: Rotating field with dephasing noise}--- Theorem~\ref{thm::main1} shows when subspace twirling can decouple system from environment, while Theorem~\ref{thm::main2} separately provides a condition for the achievability of HS/SHS via subspace twirling. The two results are not mutually exclusive so it is possible to apply both and simultaneously suppress noise and achieve SHS without adaptive controls. We demonstrate this by considering the example of a qubit in a rotating field subject to dephasing noise. 

Consider a qubit system with signal Hamiltonian \begin{align} \label{eq::qubitSignal} H_{ S}(\theta,t)=\sin(\theta t)\sigma_x+\cos(\theta t)\sigma_z.\end{align} This describes a rotating magnetic field in the $XZ$-plane with frequency $\theta$, and the goal is to estimate $\theta$. We assume the qubit probe is subject to dephasing noise, described by the system-bath interaction term \begin{align} \label{eq::qubitNoise}
	H_{\rm SB}(t)= S\otimes B,
\end{align}  where $S \coloneqq\sqrt{\epsilon(t)}\,\sigma_z $ is the noise operator acting on the probe system, $\sqrt{\epsilon(t)}$ is a time dependent dephasing rate and $B$ is some bath operator.

Let $ \ket{\pm}$ be the usual eigenbasis of $\sigma_x$ with eigenvalues $\pm1$. We can directly check that $\braket{+| B| +} = 0$, which is Condition (a) in Theorem~\ref{thm::main1}. We also have $\braket{+| H_S(\theta,t)| +} = \sin(\theta t)$, which is not a constant function, so Condition (b) in Theorem~\ref{thm::main1} is satisfied. Applying Theorem~\ref{thm::main1}, we are guaranteed that the noise can be suppressed while encoding $\theta$, and the subspace twirl that achieves the noise suppression can be chosen to be  $\mathcal{T}_{\mathcal{V}}$ where $\mathcal{V} = \operatorname{span} \{\ketbra{-}{-} \}$ (see Theorem~\ref{thm::main1}, proof of sufficiency, in \cite{SuppMa}).

Next, consider $K(\theta,t) = \int_{0}^{t} \partial_{\theta} H_S(\theta,s) \, ds$. Suppose we are estimating a low frequency $\theta$, so we are primarily interested in the measurement sensitivity locally around $\theta = 0.$ At $\theta = 0$, we have \begin{align}
	K(\theta, t) |_{\theta=0}  = \left. \left[ \int_{0}^{t} s \cos(\theta s) \sigma_x - s \sin(\theta s) \sigma_y  \, ds \right] \, \right|_{\theta = 0}  = \frac{t^2}{2} \sigma_x.
\end{align} Evaluating the operator norm we get $\Vert K(\theta, t) \Vert |_{\theta=0} = t^2/2=\Omega(t^2)$, which is the condition in Theorem~\ref{thm::main2}. This guarantees that $\mathcal{F}(\theta) = \Omega(t^4)$ is achievable via a twirling map $\mathcal{T} = \mathcal{T}_{\mathcal{V}}\circ \mathcal{T}_{\mathcal{V}^\perp}$, where $\mathcal{V} =\operatorname{span} \{\ketbra{-}{-} \}$ (see Theorem~\ref{thm::main2}, proof of sufficiency, in \cite{SuppMa}). 

For a two dimensional Hilbert space, we may write $\mathcal{T} = \mathcal{T}_{\mathcal{V}}\circ \mathcal{T}_{\mathcal{V}^\perp} = \mathcal{T}_{\mathcal{V}}$. We observe that this is the same twirling operation that decouples the system from environment. This allows a single twirling operation to simultaneously satisfy Theorems\ref{thm::main1} and~\ref{thm::main2} as claimed. The achieved scaling matches previously known bounds for AC magnetometry \cite{pang2017optimal, schmitt2017submillihertz, dey2025quantum}. The main contribution of this work is in the simultaneous suppression of noise, and the achievement of SHS without adaptive controls.

To illustrate these results, we numerically simulate the dynamics generated by the total Hamiltonian $H_{\rm tot}(\theta,t)$ in Eq.~\ref{eq::H_tot}, where the signal Hamiltonian $H_S(\theta,t)$ is given by Eq.~\ref{eq::qubitSignal}, and system-bath interaction $H_{SB}(t)$ is given by Eq.~\ref{eq::qubitNoise}. We assume that the bath is a single bosonic mode, with bath Hamiltonian $H_B = \omega_b \, a^\dagger a $, frequency $\omega_b = 2$, bath operator $B = a+a^\dagger$, and constant dephasing rate $\epsilon(t) = 0.5$. For the initial probe state, we choose $\ket{\psi_0} = (\ket{+} + \ket{-})/\sqrt{2}, $ where $\ket{\pm}$ are the eigenstates of $\sigma_x$. For the initial bath state, we choose a truncated thermal state $\rho_B \propto e^{-\beta H_B}$ with mean energy $\bar{n} = 0.5$ and Fock state cutoff at $N=6$.

We consider a finite number of control cycles $n$ over the interactions times $t \in [0,10 ]$, where the twirling map $\mathcal{T}_{\mathcal{V}}$ is applied at equally spaced intervals over a finite interaction time $t \in [0,10]$. Within each interval, the duration of time spent where the Hamiltonian is twirled chosen such that $f=1-1/n$ is the total fraction of time spent under twirling, while $1-f$ is the fraction of time spent not under any control. Under these assumptions, we see that as $n \rightarrow \infty$ the system eventually approaches a state of constant twirling, which is exactly the bang-bang control limit. When $n \rightarrow 0,$ the system recovers the original noisy dynamics described by $H_{\rm tot}(\theta,t)$ with no controls applied.

\begin{figure}[h]
		\centering
		\includegraphics[width=\linewidth]{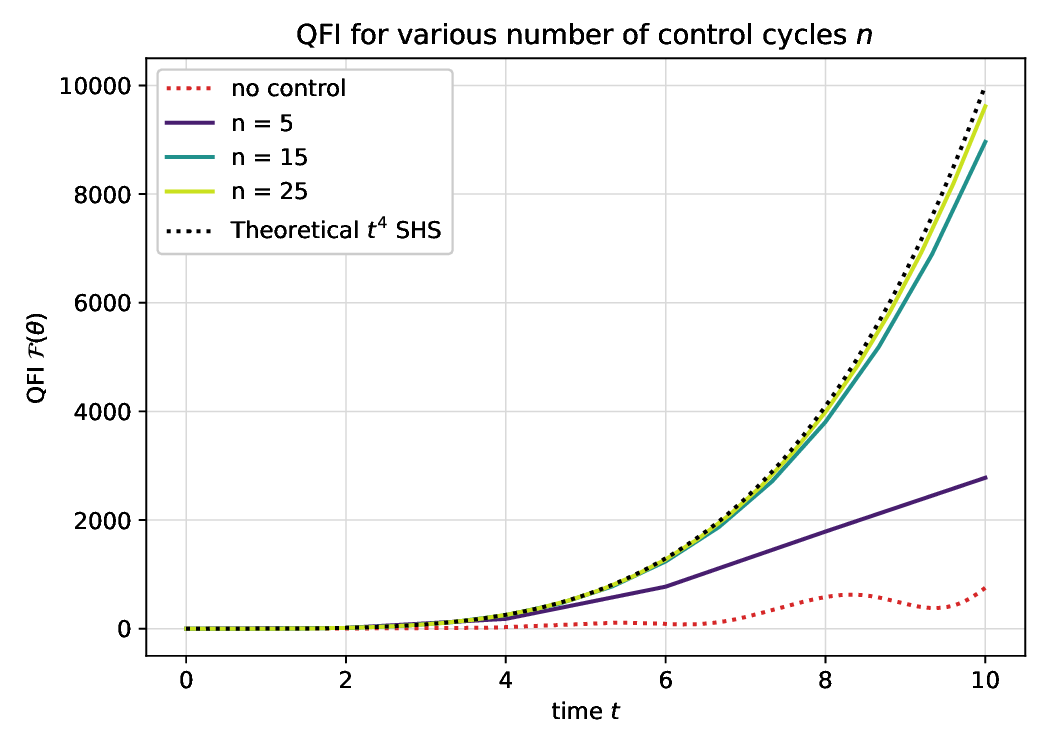}
		\caption{Time dependence of the QFI for qubit in a rotating magnatic fields subject to noise. The upper dotted line is the theoretical upper bound, while the lower dotted line is the noisy QFI without additional controls. Solid lines represent the achievable QFI for various number of control cycles $n$ applied.} \label{Fig_QFI}
	\end{figure}

In Fig.~\ref{Fig_QFI}, we plot the QFI scaling with time for various number of control cycles $n$ at the local point $\theta = 0$.  The upper dotted line represents the theoretical SHS upper limit for the QFI $\mathcal{F}(\theta) \, |_{\theta = 0}= t^4$ when there is no noise. The lower dotted line represents the lower limit of QFI where the probe is subject to noise, and no controls are applied. Observe that as the number of control cycles $n$ increases, we eventually approach the theoretical upper bound $t^4$. This is consistent with our prior prediction that subspace twirling $\mathcal{T}_\mathcal{V}$ can suppress noise and achieve SHS with scaling $\Omega(t^4).$ This control is nonadaptive, since the same control was applied at every time interval.

	%%%%%%%%%%%%%%%%%%%%%%%%%%%%%%%%%%%%%%%%%%%%%%%%%%%%%%%%%%%%%%%%%%%%%%%%%%%%%%%%%%%%%%%%%%%%%%%%%%%%%%%%%%%%%%%%%%%%%%%%%%%%%%%%%%%%%%%%%%%%%%%%%%%%%%%%%%%%%%%%%%%%%%%%%%%%%%%%%%%%%%%%%%%%
	
\textit{Discussion}---We developed a framework for achieving HS/SHS under noisy conditions in general time dependent quantum metrology,  without the use of adaptive controls. The primary tool used to achieve this is the subspace twirling operator $\mathcal{T}_{\mathcal{V}}$, which performs random unitary operations on the subspace $\mathcal{V}$ but acts trivially outside of this subspace. In Theorem~\ref{thm::main1} we provide a complete characterization of when it is possible to decouple system from environment by subspace twirling alone. In Theorem~\ref{thm::main2} we provide a characterization of when it is possible to achieve HS/SHS by using subspace twirling, and also provide performance guarantees for the achievable QFI using such a method. 

We stress that both Theorems~\ref{thm::main1} and~\ref{thm::main2} apply under very general conditions, in the sense that we have considered any time dependent signal Hamiltonian $H_S(\theta,t)$ and system-bath interaction $H_{SB}(t)$. To achieve these results, we have assumed that we are operating in the bang-bang control limit where instantaneous unitary operations can be applied to the system. We have also assumed that the parameter $\theta$ to be estimated in contained entirely within the signal Hamiltonian $H_S(\theta,t)$.

We illustrate our results with the example of a qubit probe in a rotating field subject to dephasing noise. In this example, we demonstrate the \emph{simultaneous} achievability of noise suppression and SHS without the use of adaptive controls. This is done by twirling over a single subspace $\mathcal{V} \subset \mathcal{H}_S$. In Fig.~\ref{Fig_QFI}, we see  that when the interaction times $t$ are sufficiently large relative to the number of control cycles $n$, the measurement sensitivity drops as we eventually deviate from the bang-bang limit, and Theorems~\ref{thm::main1} and~\ref{thm::main2} no longer applies. Our results therefore apply when the number of control cycles are sufficiently large relative to the interaction times $t$ considered.

Our findings show that subspace twirling is a flexible and powerful method to preserve time-dependent signals under noisy conditions, while simplifying the overall control scheme necessary to achieve a quantum advantage. This work complements recent decoupling-based strategies for periodically driven Hamiltonians~\cite{Wei2026}. Our results can help guide the design of experimental protocols in solid-state and atomic platforms, where time-dependent control fields and fast pulsed sequences are readily available~\cite{Gao2026,Yang2024}. The theoretical framework also connects directly to ongoing efforts to extend control-enhanced protocols to noisy, time-dependent signal estimation~\cite{Wan2022,previdi2026precision}. In future work, we hope to investigate other nonadaptive protocols beyond subspace twirling that may also enable SHS to be achieved under noisy conditions. Such protocols may offer new possibilities for further reducing the resource requirements for achieving SHS by relaxing the bang-bang control limit assumption.

\textit{Acknowledgments}---K.C.T. acknowledges support by the National Natural Science Fund of China and the University of Electronic Science and Technology of China.

	%%%%%%%%%%%%%%%%%%%%%%%%%%%%%%%%%%%%%%%%%%%%%%%%%%%%%%%%%%%%%%%%%%%%%%%%%%%%%%%%%%%%%%%%%%%%%%%%%%%%%%%%%%%%%%%%%%%%%%
	%%%%%%%%%%%%%%%%%%%%%%%%%%%%%%%%%%%%%%%%%%%%%%%%%%%%%%%%%%%%%%%%%%%%%%%%%%%%%%%%%%%%%%%%%%%%%%%%%%%%%%%%%%%%%%%%%%%%%%%%%%%%%%%%%%%%%
	\FloatBarrier
	%%%%%%%%%%%%%%%%%%%%%%%%%%%%%%%%%%%%%%%%%%%%%%%%%%%%%%%%%%%%%%%%%%%%%%%%%%%%%%%%%%%%%%%%%%%%%%%%%%%%%%%%%%%%%%%%%%%%%%%%%%%%%%%%%%%%%%%%%%%%%%%%%%%%%%%%%%%%%%%%%%%%%%%%%%%%%%%%%%%%%%%%%%%%%%%%%%%%%%%%%%%%%%%%%%%%%%%%%%%%%%%%

	%%%%%%%%%%%%%%%%%%%%%%%%%%%%%%%%%%%%%%%%%%%%%%%%%%%%%%%%%%%%%%%%%%%%%%%%%%%%%%%%%%%%%%%%%%%%%%%%%%%%%%%%%%%%%%%%%%%%%%%%%%%%%%%%%%%%%%%%%%%%%%%%%%%%%%%%%%%%%%%%%%%%%%%%%%%%%%%%%%%%%%%%%%%%%%%%%%%%%%%%%%%%%%%%%%%%%%%%%%%%%%%%%%%%%%%%%%%%%%%%%%%%%%%%%%%
	
	\newpage
	\bibliography{references}
	
	%%%%%%%%%%%%%%%%%%%%%%%%%%%%%%%%%%%%%%%%%%%%%%%%%%%%%%%%%%%%%%%%%%%%%%%%%%%%%%%%%%%%%%%%%%%%%%%%%%%%%%%%%%%%%%%%%%%%%%%%%%%%%%%%%%%%%%%%%%%%%%%%%%%%%%%%%%%%%%%%%%%%%%%%%%%%%%%%%%%%%%%%%%%%%%%%%%%%%%%%%%%%%%%%%%%%%%%%%%%%%%%%%%%%%%%%%%%%%%%%%%%%%%%%%%%%%%%%%%%%%%%%%%%%%%
	\onecolumngrid
	
\end{document}

\title{Supplemental material for ``Noise robust super-Heisenberg scaling in time-dependent quantum sensing by subspace-twirled dynamical decoupling"}

\author{Bakmou Lahcen}
\affiliation{ Institute of Fundamental and Frontier Sciences,
University of Electronic Science and Technology of China, Chengdu 610051, China}
\affiliation{Key Laboratory of Quantum Physics and Photonic Quantum Information, Ministry of Education, University of Electronic Science and Technology of China, Chengdu 611731, China}

\author{Ke Zeng}
\affiliation{ Institute of Fundamental and Frontier Sciences,
University of Electronic Science and Technology of China, Chengdu 610051, China}
\affiliation{Key Laboratory of Quantum Physics and Photonic Quantum Information, Ministry of Education, University of Electronic Science and Technology of China, Chengdu 611731, China}

\author{Tianyu Yang}
\affiliation{ Institute of Fundamental and Frontier Sciences,
University of Electronic Science and Technology of China, Chengdu 610051, China}
\affiliation{Key Laboratory of Quantum Physics and Photonic Quantum Information, Ministry of Education, University of Electronic Science and Technology of China, Chengdu 611731, China}

\author{Yu Jiang}
\affiliation{ Institute of Fundamental and Frontier Sciences,
University of Electronic Science and Technology of China, Chengdu 610051, China}
\affiliation{Key Laboratory of Quantum Physics and Photonic Quantum Information, Ministry of Education, University of Electronic Science and Technology of China, Chengdu 611731, China}

\author{Kok Chuan Tan}
\affiliation{ Institute of Fundamental and Frontier Sciences, University of Electronic Science and Technology of China, Chengdu 610051, China}
\affiliation{Key Laboratory of Quantum Physics and Photonic Quantum Information, Ministry of Education, University of Electronic Science and Technology of China, Chengdu 611731, China}
\affiliation{School of Physics, University of Electronic Science and Technology of China, Chengdu 610054, China}
\affiliation{School of Science, Xihua University, Chengdu 610039, China}
\email{tankokchuanbobby@uestc.edu.cn}
%%%%%%%%%%%%%%%%%%%%%%%%%%%%%%%%%%%%%%%%%%%%%%%%%%%%%%%%%%%%%%%%%%%%%%%%%%%%%%%%%%%%%%%%%%%%%%%%%%%%%%%%%%%%%%%%%%%%%%%%%%%%%%%%%%%%%%%%%%%%%%%%%%%%%%%%%%%%%%%%%%%%%%%%%%%%%%%%%%%%%%%%%%%%%%%%%%%%%%%%%%%%%%%%%%%%%%%%%%%%%%%%%%%%%%%%%%%%%%%%%%%%%%%%%%%

\maketitle

%%%%%%%%%%%%%%%%%%%%%%%%%%%%%%%%%%%%%%%%%%%%%%%%%%%%%%%%%%%%%%%%%%%%%%%%%%%%%%%%%%%%%%%%%%%%%%%%%%%%%%%%%%%%%%%%%%%%%%%%%%%%%%%%%%%%%%%%%%%%%%%%%%%%%%%%%%%%%%%%%%%%%%%%%%%%%%%%%%%%%%%%%%%%%%%%%%%%%%%%%%%%%%%%%%%%%%%%%%%%%%%%%%%%%%%%%%%%%%%%%%%%%%%%%%%%%%%%%%%%%%%%%%%%%%%%%%%%%%%%%%%%%%%%%%%%%%%%%%%%%%%%%%%%%%%%%%%%%%%%%%%%%%%%%%%%%%%%%%%%%%%%%%%%%%%%%%%%%%%%%%%%%%%%%%%%%%%%%%%%%%%%%%%%%%%%%%%%%%%%%%%%%%%%%%%%%%%%%%%%%%%%%%%%%%%%%%%%%%%%%%%%%%%%%%%%%%%%%%%%%%%%%%%%%%%%%%%%%%%%%%%%%%%%%%%%%%%%%%%%%%%%%%%%%%%%%%%%%%%%%%%%%%%%%%%%%%%%%%%%%%%%%%%%%%%%%%%%%%%%%%%%%%%%%%%%%%%%%%%%%%%%%%%%%%%%%%%%%%%%%%%%%%%%%%%%%%%%%%%%%%%%%%%%%%%%%%%%%%%%%%%%%%%%%%%%%%%%%%%%%%%%%%%%%%%%%%%%%%%%%%%%%%%%%%%%%%%%%%%%%%%%%%%%%%%%%%%%%%%%%%%%%%%%%%%%%%%%%%%%%%%%%%%%%%%%%%%%%%%%%%%%%%%%%%%%%%%%%%%%%%%%%%%%%%%%%%%%%%%%%%%%%%%%%%%%%%%%%%%%%%%%%%%%%%%%%%%%%%%%%%%%%%%%%%%%%%%%%%%%%%%%%%%%%%%%%%%%%%%%%%%%%%%%%%%%%%%%%%%%%%%%%%%%%%%%%%%%%%%%

\section{Necessary and sufficient conditions for noise suppression and signal preservation using subspace twirling}

In this section, we present a complete proof of Theorem 1 in the main text. We will consider a general time-dependent Hamiltonian of the following form: 
\begin{equation} \label{eq::H_tot}
	H_{\rm {tot}}(\theta, t) = H_{\rm S}(\theta, t) \otimes \openone_{\rm B} + \openone_{\rm S} \otimes H_{\rm B}(t) + H_{\rm SB}(t),
\end{equation}
where $H_{\rm S}(\theta,t) = \sum\nolimits_{i=1}^{r} g_i(\theta,t)\, G_i$ is the signal Hamiltonian, $G_i$ are Hermitian operators, $g_i(\theta,t)$ are time-dependent functions encoding unknown parameter $\theta$, and $H_{\rm SB}(t) = \sum_j  h_j(t) \, S_{j}\otimes B_{j}$ is a generic time dependent, system-bath interaction in some operator basis $\{S_i\}$ and $\{ B_j \}$.The goal is to find a strict subspace $\mathcal{V} \in \mathcal{H}_S$ and corresponding subspace twirl $\mathcal{T}_{\mathcal{V}}$ that satisfies the following:

\begin{align} \label{eq::decouple}
	\mathcal{T}_{\mathcal{V}} \otimes \openone [H_{\rm SB}(t)] &\propto \openone_S \otimes B'(t), \\
	 \label{eq::signalCondition1}
	\mathcal{T}_{\mathcal{V}} [H_S(\theta,t)] &\not\propto H_S'(t), \\ \label{eq::signalCondition2}
	\mathcal{T}_{\mathcal{V}} [H_S(\theta,t)]  &\not\propto \openone_S.
\end{align}

Eq.~\ref{eq::decouple} is the requirement that the subspace twirl $\mathcal{T}_{\mathcal{V}}$ decouples the system from environment, while Eqs.~\ref{eq::signalCondition1} and~\ref{eq::signalCondition2} is the requirement that a nontrivial encoding of parameter $\theta$ survives the subspace twirling process.

The statement to be proven is the following.
\begin{thm} \label{thm::main1}
	Given $H_{\rm tot}(\theta,t)$, there exists a strict subspace $\mathcal{V} \in \mathcal{H}_S$ and corresponding subspace twirling $\mathcal{T}_{\mathcal{V}}$ that satisfies the noise suppression and signal preserving conditions  Eqs.~\ref{eq::decouple}, \ref{eq::signalCondition1} and~\ref{eq::signalCondition2} if and only if there exists a vector $\ket{v} \in \mathcal{H}_S$ such that (a) $\braket{v | S_i |v} = 0$ for every $i$ and (b) $\nu(\theta) \coloneqq \braket{v | H_S(\theta, t) |v} $ is not a constant function with respect to $\theta$.
\end{thm}

\begin{proof}
	\noindent\textbf{(Proof of necessity)} Suppose there exists $\mathcal{V} \in \mathcal{H}_S$ and $\mathcal{T}_{\mathcal{V}}$ that satisfies Eqs.~\ref{eq::decouple}, \ref{eq::signalCondition1} and~\ref{eq::signalCondition2}. Without any loss in generality, we can assume that $S_i$ are traceless Hermitian operators. We can make this assumption because if $S_i$ has nonzero trace,  it contributes a Hamiltonian term proportional to $\openone \otimes B_j$. Such terms act nontrivially only on the bath $\mathcal{B}$, so they can always be grouped with the $\openone_S \otimes H_B(t)$. This allows us to always write $H_{\rm SB}(t) = \sum_j  h_j(t) \, S_{j}\otimes B_{j}$ where $S_j$ is a traceless operator.

	Condition Eqs.~\ref{eq::decouple} implies that $\sum_{j} \mathcal{T}_{\mathcal{V}}(S_j) \otimes B_j \propto \openone \otimes B'(t)$, which is equivalent to requiring $\mathcal{T}_{\mathcal{V}}(S_i) \propto \openone$. Since $S_i$ is traceless, and $\mathcal{T}_{\mathcal{V}}$ is a trace preserving map, this is equivalent to the condition \begin{align}
		\mathcal{T}_{\mathcal{V}}(S_i) = 0, \quad \forall i. \label{eq::decouple2}
	\end{align} Recall the definition of the twirling map: \begin{align} 
		\mathcal{T}_{\mathcal{V}}(X) \coloneqq \int  U X U^\dagger d\mu_H(U),
	\end{align} where $\mu_H$ is the Haar measure over random unitaries $U$ acting on the subspace $\mathcal{V}$. This means that $U$ acts trivially on the orthogonal subspace $\mathcal{V}^\perp$, but nontrivially on $\mathcal{V}$. Let $d\coloneqq \dim \mathcal{H}_S$,   $\{ \ket{e_i} \}_{i=1}^r$ be an orthonormal basis for $\mathcal{V}$, and $\{ \ket{e_i} \}_{i=r+1}^d$ be an orthonormal basis for $\mathcal{V}^\perp$. Then in the basis $\{ \ket{e_i} \}_{i=1}^d$, we can write $S_i$ and $\mathcal{T}_{\mathcal{V}}(S_i)$ in block matrix form:
	\begin{align}
		S_i &= \begin{pmatrix}
			A_i & B_i \\
			C_i & D_i
		\end{pmatrix}, \\
		\mathcal{T}_{\mathcal{V}}(S_i) &=\begin{pmatrix}
			\frac{\Tr(A_i)}{r} \openone_{\mathcal{V}} & 0 \\
			0 & D_i.
		\end{pmatrix}
	\end{align} To satisfy Eq.~\ref{eq::decouple} and hence Eq.~\ref{eq::decouple2}, we must have $D_i = 0$ for every $i$. This implies that for any $\ket{v} \in \mathcal{V}^\perp$, we satisfy \begin{align}
		\braket{v | S_i |v} = 0, \quad \forall i,
	\end{align} which satisfies Condition (a) of the Theorem. 
	
	Similarly, using the fact that $H_S(\theta,t)$ is traceless and applying the subspace twirl $\mathcal{T}_{\mathcal{V}}$ , we obtain in block matrix form \begin{align}
		\mathcal{T}_{\mathcal{V}}[H_S(\theta,t)] &= \begin{pmatrix}
		-\frac{\Tr \left [ E(\theta,t) \right ]}{r} \openone_\mathcal{V} & 0 \\
		0 & E(\theta,t),
	\end{pmatrix}, 
	\end{align} where we note that the submatrix $E(\theta,t)$ must be Hermitian since $\mathcal{T}_{\mathcal{V}}[H_S(\theta,t)]$ is Hermitian. Eq.~\ref{eq::signalCondition1} implies that $E(\theta,t)$ is not constant with respect to $\theta$ while Eq.~\ref{eq::signalCondition2} implies that $E(\theta,t) \neq -\frac{\Tr \left [ E(\theta,t) \right ]}{r} \openone_{\mathcal{V}^\perp}$. 
	
	Expand $E(\theta,t)$ in the basis $\{ \ket{e_i}\}_{i=r+1}^{d}$ such that $E(\theta,t) = \sum_{ij} E_{ij}(\theta,t) \ketbra{e_i}{e_j} $. Since $E(\theta, t)$ is not constant with respect to $\theta$, there exists at least one matrix component $\ketbra{e_i}{e_j}$ whose coefficient $ E_{ij}(\theta,t)$ is not a constant function. Suppose there exists some $i$ such that $E_{ii}(\theta, t)$ is not a constant function, then we can choose $\ket{v} = \ket{e_i} \in \mathcal{V}^\perp$ such that $\braket{v|H_S(\theta,t)|v} = \braket{v|E(\theta,t)|v} = E_{ii}(\theta,t)$ is not a constant function.  
	
	Suppose instead that $E_{ii}(\theta, t) = E'_{ii}(t)$ are all constant functions of $\theta$ for $i= (r+1),\ldots, d$. Then there must be some $i \neq j$ where $E_{ij}(\theta, t)$ is not a constant function. Allowing $E(\theta,t)$ to be complex, this implies that either (1) $\real \left[ E_{ij}(\theta, t) \right]$, and/or (2) $\imaginary \left[ E_{ij}(\theta, t) \right]$ is not a constant function of $\theta$.
	
	If (1) is true, we choose $\ket{v} = (\ket{e_i} + \ket{e_j})/\sqrt{2} \in \mathcal{V}^\perp$. One may verify that for this choice $\braket{v|H_S(\theta,t)|v}  = \braket{v|E(\theta,t)|v} = E'_{ii}(t)+E'_{jj}(t) + 2 \real \left[ E_{ij}(\theta, t) \right],$ which is not a constant function since we assumed $\real \left[ E_{ij}(\theta, t) \right]$ is not a constant function.
	
	If (2) is true, we choose $\ket{v} = (\ket{e_i} + i \ket{e_j})/\sqrt{2} \in \mathcal{V}^\perp$. One may verify that for this choice $\braket{v|H_S(\theta,t)|v}  = \braket{v|E(\theta,t)|v} = E'_{ii}(t)+E'_{jj}(t) + 2 \imaginary \left[ E_{ij}(\theta, t) \right],$ which is not a constant function since we assumed $\imaginary \left[ E_{ij}(\theta, t) \right]$ is not a constant function.
	
	We therefore conclude that in call cases, one may always find some vector $\ket{v}$ such that $\braket{v|H_S(\theta,t)|v}$ is not a constant function. This satisfies Condition (b) of the Theorem. Since both Conditions (a) and (b) can be satisfied simultaneously, this proves necessity. 
	
	\noindent\textbf{(Proof of sufficiency)} Suppose there exists a vector $\ket{v} \in \mathcal{H}_S$ such that (a) $\braket{v | S_i |v} = 0$ for every $i$ and (b) $\nu(\theta) \coloneqq \braket{v | H_S(\theta, t) |v} $ is not a constant function with respect to $\theta$. Let $\mathcal{V}^\perp \coloneqq \operatorname{span}\{\ket{v}\}$ be a one dimensional subspace of $\mathcal{H}_S$, and $\mathcal{V}$ be the $(d-1)$ dimensional orthogonal subspace, where $d\coloneqq \dim \mathcal{H}_S$. We then choose the twirling map $\mathcal{T}_{\mathcal{V}}$. Applying it to $S_i, $ we verify that in block matrix form \begin{align}
		\mathcal{T}_{\mathcal{V}} (S_i) = \begin{pmatrix}\frac{-\braket{v|S_i|v}}{d-1} \openone_{\mathcal{V}} & 0 \\
			0 & \braket{v|S_i|v}
		\end{pmatrix} .
	\end{align} Since we assumed Condition (a) is true, this implies that \begin{align}
	\mathcal{T}_{\mathcal{V}} (S_i) = 0, \quad \forall i,
	\end{align} which we have already shown in equivalent to Eq.~\ref{eq::decouple}.
	
	We can also verify that \begin{align} \label{eq::blockMatrixHS}
		\mathcal{T}_{\mathcal{V}} (H_S(\theta,t)) = \begin{pmatrix}\frac{-\nu(\theta,t)}{d-1} \openone_{\mathcal{V}} & 0 \\
			0 & \nu(\theta,t)
		\end{pmatrix} .
	\end{align} Since we assumed Condition (b) is true, $v(\theta, t)$ is not a constant function of $\theta$, hence neither is $\mathcal{T}_{\mathcal{V}} (H_S(\theta,t))$. This implies Eq.~\ref{eq::signalCondition1} is satisfied.
	
	Finally, if $\mathcal{T}_{\mathcal{V}} [H_S(\theta,t)]  \propto \openone_S,$, the diagonal entries in Eq.~\ref{eq::blockMatrixHS} must be equal to each other, so $-\frac{v(\theta,t)}{d-1} = v(\theta,t)$. Since $d-1 \geq 1$, this can only be true if $v(\theta,t) = 0$. However, this is a constant function, which contradicts the assumption that Condition (b) is true. Therefore, we must have that $\mathcal{T}_{\mathcal{V}} [H_S(\theta,t)]  \not\propto \openone_S,$ which is Eq.~\ref{eq::signalCondition2}. We have therefore shown that if Conditions (a) and (b) of the Theorem is satisfied, then there exists at least one strict subspace $\mathcal{V} \subset \mathcal{H}_S$ and corresponding twirling map $\mathcal{T}_{\mathcal{V}}$ that satisfies Eqs.~\ref{eq::decouple}, \ref{eq::signalCondition1} and~\ref{eq::signalCondition2}. This proves sufficiency.

\end{proof}

\section{Example: Noise suppression for specific probe states}

In this section, we discuss an explicit example where the system-bath dynamics is not completely decoupled in the sense of Eq.~\ref{eq::decouple}, but a specific probe state $\ket{\psi_0}$ may nevertheless experience noiseless dynamics through subspace twirling. We demonstrate how Theorem~\ref{thm::main1} may be applied in such cases through the following example.

Consider a qutrit system where $\mathcal{H}_S = \operatorname{span} \{ \ket{0}, \ket{1}, \ket{2}\}$. For the signal Hamiltonian, we choose \begin{align}
    H_S(\theta, t) \coloneqq  \theta t \begin{pmatrix}
        1 & 0 & 0 \\
        0 & -1 & 0 \\
        0 & 0 & 0
    \end{pmatrix}.
\end{align} For the system-bath interaction $H_{\rm SB}(t)$ we choose $H_{\rm SB}(t) \coloneqq \sum_{i=1}^3 S_i \otimes B_i,$ where \begin{align}
    S_1 \coloneqq \begin{pmatrix}
        1 & 1/2 & 0 \\
        1/2 & 1 & 0 \\
        0 & 0 & -2
    \end{pmatrix}, \quad S_2 \coloneqq \begin{pmatrix}
        0 & 0 & 1 \\
        0 & 0 & 1 \\
        1 & 1 & 0
    \end{pmatrix},  \quad S_3 \coloneqq \begin{pmatrix}
        0 & 0 & -i \\
        0 & 0 & -i \\
        i & i & 0
    \end{pmatrix},
\end{align} and $B_j$ can be any bath operator. Suppose the initial probe state is $\ket{\psi_0} = (\ket{0} + \ket{1})/\sqrt{2}$. Observe that the \emph{noiseless} dynamics generated by $H_S(\theta,t)$ ensures that $U(\theta,t) \ket{\psi_0} \in \mathcal{W}$, where $\mathcal{W} = \operatorname{span} \{ \ket{0}, \ket{1} \}$ is a two dimensional subspace of $\mathcal{H}_S$. Therefore, for this probe state, we don't necessarily need a full decoupling of the system and bath. Rather, we just need to ensure that the system and bath is effectively decoupled within the subspace $\mathcal{W}$. We will show how to apply Theorem~\ref{thm::main1} to this situation.

The idea is to restrict the Hamiltonian to the subspace $\mathcal{W}$ before applying Theorem~\ref{thm::main1}. This is achieved by defining the projector $\Pi_{\mathcal{W}} = \ketbra{0}{0} + \ketbra{1}{1}$ onto the subspace $\mathcal{W}$, and then projecting onto the subspace $\mathcal{W}$ by performing the replacements $ H_{\rm  SB}(t) \rightarrow ( \Pi_{\mathcal{W}}\otimes \openone) H_{\rm  SB} (t) (\Pi_{\mathcal{W}}\otimes \openone)$,  $H_S(\theta,t) \rightarrow \Pi_{\mathcal{W}} H_S(\theta,t) \Pi_{\mathcal{W}}$, and $\openone \rightarrow \Pi_{\mathcal{W}}$ in Eqs.~\ref{eq::decouple}, \ref{eq::signalCondition1} and~\ref{eq::signalCondition2}, which is the recipe described in the main text. After projection, we obtain the following: 
\begin{align}
     \Pi_{\mathcal{W}} H_S(\theta,t) \Pi_{\mathcal{W}}  =  \theta t \begin{pmatrix}
        1 & 0 & 0 \\
        0 & -1 & 0 \\
        0 & 0 & 0
    \end{pmatrix},
\end{align} and \begin{align}
      \Pi_{\mathcal{W}} S_1  \Pi_{\mathcal{W}} = \begin{pmatrix}
        1 & 1/2 & 0 \\
        1/2 & 1 & 0 \\
        0 & 0 & 0
    \end{pmatrix}, \quad    \Pi_{\mathcal{W}} S_2  \Pi_{\mathcal{W}} = \begin{pmatrix}
        0 & 0 & 0 \\
        0 & 0 & 0 \\
        0 & 0 & 0
    \end{pmatrix},  \quad    \Pi_{\mathcal{W}} S_3  \Pi_{\mathcal{W}} = \begin{pmatrix}
        0 & 0 & 0 \\
        0 & 0 & 0 \\
        0 & 0 & 0
    \end{pmatrix}. 
\end{align} We see that $\Pi_{\mathcal{W}} S_2  \Pi_{\mathcal{W}}= \Pi_{\mathcal{W}} S_3  \Pi_{\mathcal{W}} = 0 $, so they no longer contribute noise to the subspace $\mathcal{W}$. On $\mathcal{W}$, the effective dynamic is described by the operators \begin{align}
    H'_S(\theta,t) \coloneqq \theta t\, \sigma_z,\quad S_1' \coloneqq \sigma_x/2,
\end{align} where $\sigma_z$ and $\sigma_x$ are the usual Pauli operators acting on $\mathcal{W}$. In particular for $S'_1$, we have removed the identity component that contributes a nonzero trace and generates a trivial type of noise on $\mathcal{W}$. Applying Theorem~\ref{thm::main1} to $H'_S(\theta,t)$ and $S_1'$, we see that $\braket{0 | S'_1 |0 } = 0$ and $\braket{0 | H'_S(\theta,t) |0 } = \theta t$ , so the Conditions (a) and (b) of the Theorem are satisfied. This implies that subspace twirling can achieve noiseless dynamics for the probe state $\ket{\psi_0} \in \mathcal{W}$, while nontrivial encoding of $\theta$ onto $\ket{\psi_0}$ occurs. 

As for the choice of the twirled subspaces, we choose $\mathcal{T} = \mathcal{T}_{\mathcal{V}_1} \circ \mathcal{T}_{\mathcal{V}_2}$ where $\mathcal{V}_1 = \operatorname{span} \{ \ket{1}\}$ and $\mathcal{V}_2 = \operatorname{span} \{\ket{2}\}$. Here, $\mathcal{V}_1$ is chosen because it is the subspace of $\mathcal{W}$ that needs to be twirled to suppress noise on $\mathcal{W}$. This comes from Theorem~\ref{thm::main1}. $\mathcal{V}_2$ is chosen because it defines the orthogonal subspace $\mathcal{W}^\perp$, which is the part of the Hilbert space projected away when we restricted the Hamiltonian to $\mathcal{W}$. We can verify that \begin{align} \label{eq::encoding}
     \mathcal{T} [ H_S(\theta,t) ]  =  \theta t \begin{pmatrix}
        1 & 0 & 0 \\
        0 & -1 & 0 \\
        0 & 0 & 0
    \end{pmatrix},
\end{align} and \begin{align} \label{eq::noiseOp}
      \mathcal{T} [ S_1 ]= \begin{pmatrix}
        1 & 0 & 0 \\
        0 & 1 & 0 \\
        0 & 0 & -2
    \end{pmatrix}, \quad    \mathcal{T} [ S_2 ] = \mathcal{T} [ S_2 ]  = \begin{pmatrix}
        0 & 0 & 0 \\
        0 & 0 & 0 \\
        0 & 0 & 0
    \end{pmatrix}.
\end{align} Indeed, we see that the twirled Hamiltonian Eq.~\ref{eq::encoding} nontrivially encodes $\theta$ onto the probe state $\ket{\psi_0} = (\ket{0} + \ket{1})/\sqrt{2}$, while the noise terms Eq.~\ref{eq::noiseOp} acts trivially on any vector $a_1\ket{0}+a_2\ket{1}$ from the subspace $\mathcal{W}$.

\section{Necessary and sufficient conditions for Heisenberg/super-Heisenberg scaling without adaptive controls using subspace twirling}

In this section, we present a complete proof of Theorem 2 in the main text. In the following result, we say that HS or SHS is achieved if $\mathcal{F}(\theta) = \Omega(t^{l}) $ for some $l \geq 2 $.  The statement to be proven is the following:

\begin{thm} \label{thm::main2}
	Given $H_{\rm S}(\theta,t) = \sum\nolimits_{i=1}^{r} g_i(\theta,t)\, G_i$, there exists a strict subspace $\mathcal{V} \subset \mathcal{H}_S$ and corresponding twirling map $\mathcal{T} \coloneqq \mathcal{T}_{\mathcal{V}} \circ \mathcal{T}_{\mathcal{V}^\perp}$ such that $\mathcal{T}[H_S(\theta,t)]$ achieves HS or SHS without additional adaptive controls if and only if $\Vert K(\theta,t)\Vert = \Omega(t^k)$ for some $k \geq 1$ where $K(\theta,t) \coloneqq \int_{0}^{t} \partial_{\theta}H_S(\theta,s) \, ds $.
	
	Furthermore, if $\Vert K(\theta,t)\Vert = \Omega(t^k)$, then the QFI $\mathcal{F}(\theta) = \Omega(t^{2k})$ is achievable.
\end{thm}

\begin{proof}
	
	\noindent\textbf{(Proof of necessity)} Suppose there exists a strict subspace $\mathcal{V} \subset \mathcal{H}_S$ and corresponding twirling map $\mathcal{T} \coloneqq \mathcal{T}_{\mathcal{V}} \circ \mathcal{T}_{\mathcal{V}^\perp}$ such that $\mathcal{T}[H_S(\theta,t)]$ achieves HS or SHS without additional adaptive controls.
	
	Since $\mathcal{T}$ twirls over the orthogonal subspaces $\mathcal{V}$ and $\mathcal{V}^\perp$, this implies that in the basis of $\mathcal{V}$ and $\mathcal{V}^\perp$, we can write $ \partial_\theta \mathcal{T}[H_{\rm S}(\theta,t)]$ in block matrix form: \begin{align}
		\partial_\theta \mathcal{T}[H_{\rm S}(\theta,t)] = \begin{pmatrix}
			\frac{\beta(\theta,t)}{m} \openone_{\mathcal{V}}  & 0 \\ 0 & -\frac{\beta(\theta,t)}{d-m} \openone_{\mathcal{V}^\perp},
		\end{pmatrix}
	\end{align} where $d \coloneqq \dim \mathcal{H}_S$, $m \coloneqq \dim \mathcal{V} $, $\beta(\theta,t) \coloneqq \sum_i \beta_i(\theta,t)$, and $\beta_i(\theta,t) \coloneqq \partial_{\theta} g_i(\theta,t) \sum_{j=1}^m \braket{e_j|G_i |e_j}$. Observe that the maximum and minimum eigenvalues of $ \partial_\theta \mathcal{T}[H_{\rm S}(\theta,t)]$ are either $\frac{\beta(\theta,t)}{m}$ or $-\frac{\beta(\theta,t)}{d-m}.$ Since we have assumed that HS/SHS is achievable without adaptive controls, this implies that HS/SHS is also achievable if adaptive controls are allowed. Allowing for adaptive controls we get the following an upper bound on the (non-adaptive) maximum QFI: \begin{align}
		 \label{eq::upperBound} \max_{\ket{\psi_0}} \mathcal F(\theta) &\le \left[ \int_0^t \left( \lambda_{\max}(s) - \lambda_{\min}(s) \right) ds \right]^2 \\
		&=(\frac{1}{m}+\frac{1}{d-m})^2 \left [ \int_0^t  \beta(\theta,s) ds \right]^2 \\
		&= \Omega(t^{2k}),
	\end{align} where $k\geq 1$ since we assumed HS or SHS is achievable. This implies that $\abs{\int_0^t  \beta(\theta,t) ds} = \Omega(t^{k})$ where $k\geq 1$.
	
	Since $\beta(\theta,t) \coloneqq \sum_i \beta_i(\theta,t)$, we have the chain of inequalities\begin{align}
		\abs{\int_0^t  \beta(\theta,t) ds} &= \abs{\int_0^t  \sum_i\partial_{\theta} g_i(\theta,t) \sum_{j=1}^m \braket{e_j|G_i |e_j} ds} \\
		&= \abs{\sum_{j=1}^m \braket{e_j|K(\theta,t) |e_j}} \\
		&\leq m \Vert K(\theta,t) \Vert.
	\end{align} We obtain $\Vert K(\theta,t) \Vert \geq \Omega(t^k)/m$ which implies $\Vert K(\theta,t) \Vert = \Omega(t^k)$ where $k\geq 1$ since $m$ is a constant independent of $t$. This proves necessity.
	
	\hfill 
	
	\noindent\textbf{(Proof of sufficiency)} Assume that  $\Vert K(\theta,t) \Vert = \Omega(t^k)$ where $k\geq 1$. Applying the definition of the operator norm and observing that $K(\theta,t)$ is Hermitian, let $\ket{v}$ be a normalized vector such that $\Vert K(\theta,t) \Vert = \abs{\braket{v | K(\theta,t)|v}} = \Omega(t^k)$.
	
	Let $\mathcal{V} \coloneqq \operatorname{span}\{\ket{v}\}$ be a one dimensional subspace of $\mathcal{H}_S$, and $\mathcal{V}^\perp$ be the $(d-1)$ dimensional orthogonal subspace, where $d\coloneqq \dim \mathcal{H}_S$. Consider the twirling map $\mathcal{T} \coloneqq \mathcal{T}_{\mathcal{V}} \circ \mathcal{T}_{\mathcal{V}^\perp}$. Applying to both $H_S(\theta,t)$ and $\partial_{\theta} H_S(\theta,t)$, one may verify that both $\mathcal{T}[H_S(\theta,t)]$  and $\partial_{\theta} \mathcal{T}[ H_S(\theta,t)]$ are diagonal in the same basis, such that \begin{align}
		[\mathcal{T}[H_S(\theta,t)],\partial_{\theta} \mathcal{T}[ H_S(\theta,t)]]=0.
	\end{align}  This implies that $U(\theta,t)=\mathscr{T}\exp{-i\int_0^t \mathcal{T}[H_S(\theta,t)] ds}$ commutes with  $\partial_{\theta} \mathcal{T}[ H_S(\theta,t)]$, so the generator $h(\theta,t)$ is given by the expression \begin{align}
	h(\theta,t) &= \int_{0}^{t} U^\dagger(\theta,s) \partial_{\theta} \mathcal{T}[ H_S(\theta,s)]  U(\theta,s) \, ds \\
	&= \int_{0}^{t} \partial_{\theta} \mathcal{T}[ H_S(\theta,s)] \, ds \\
	& = \begin{pmatrix}
		\int_{0}^{t} \beta(\theta,s) \, ds   & 0 \\ 0 & -\frac{\int_{0}^{t} \beta(\theta,s) \, ds}{d-1} \openone_{\mathcal{V}^\perp}  
	\end{pmatrix},	\end{align} where $\beta(\theta,t) \coloneqq \sum_i \partial_{\theta}g_i(\theta,t) \braket{v|G_i|v}.$ Given the generator, one may always evaluate the achievable quantum Fisher information with initial probe state $\ket{\psi_0}$ via the expression $\mathcal F(\theta) = 4\,\operatorname{Var}[h(\theta,t)]_{\ket{\psi_0}}$. Let $\ket{\psi_0} \coloneqq (\ket{v}+\ket{v^\perp})/\sqrt{2}$ where $\ket{v^\perp}\in \mathcal{V}^\perp$. We obtain \begin{align}
	\mathcal F(\theta) =  \left (1+\frac{1}{d-1} \right )^2 \left (\int_{0}^{t} \beta(\theta,s)  \, ds \right)^2. 
\end{align}

Finally, observe that \begin{align}
	\abs{\int_{0}^{t} \beta(\theta,s)  \, ds} &=  \abs{\int_{0}^{t}  \sum_i \partial_{\theta}g_i(\theta,s) \braket{v|G_i|v}  \, ds}  \\
	&= \abs{ \bra{v} \int_{0}^{t}  \partial_{\theta} H_S(\theta,s) \ket{v}} \\
	&= \abs{ \bra{v} K(\theta,t) \ket{v}} \\
	&= \Vert K(\theta,t) \Vert.
\end{align} This implies that the quantum Fisher information \begin{align}
\mathcal F(\theta) = \left (1+\frac{1}{d-1} \right )^2 \Vert K(\theta,t) \Vert^2 = \Omega(t^{2k}) \label{eq::achievableQFI}
\end{align} is achievable. Furthermore, since we assumed $k\geq 1$ and we have achieved this using initial state $\ket{\psi_0}$ without applying adaptive controls, we have achieved HS/SHS without adaptive controls as required. This proves sufficiency. Eq.~\ref{eq::achievableQFI} also explicitly shows that if $\Vert K(\theta,t) \Vert = \Omega(t^k)$, then the QFI $\mathcal{F}(\theta) = \Omega(t^{2k})$ is achievable, which is the statement in the last part of the Theorem and completes the proof.
	
\end{proof} 

 %\bibliography{SMbbliography}
%%%%%%